\documentclass[twocolumn,showpacs,preprintnumbers,amsmath,amssymb]{revtex4}
\usepackage{graphicx}
\usepackage{dcolumn}
\usepackage{bm}
\begin{document}

\title{Critical Properties of an Ising Model with 
Dilute Long-range Interactions}
 
\author{R.T. Scalettar}
                 
\affiliation{
Physics Department,
University of California,
Davis, CA   95616}                       

\author{
[Published as Physica {\bf A170}, 282 (1991)]}
 
\begin{abstract}
Statistical mechanical models with local interactions in $d>1$ dimension 
can be
regarded as $d=1$ dimensional models with regular long range interactions.  
In this paper we study the critical properties of Ising models
having $V$ sites, each having $z$
randomly chosen neighbors.  For $z=2$ the model reduces to the $d=1$
Ising model.  For $z= \infty$ we get a mean field model.   
We find that for finite $z > 2$ the system has a second order phase
transition characterized by a length scale $L={\rm ln}V$ 
and mean field critical exponents that are 
independent of $z$.
\end{abstract}
\pacs{05.50, 05.70.J, 05.70.F}
\maketitle
 
                   
The dimensionality of a system and the symmetries of
its Hamiltonian determine its critical behavior,
which is characterized by
the exponents that describe the singularities
in magnetization, susceptibility, and specific heat 
at the critical temperature 
$T_{c}$.  Meanwhile,
features of a model such as additional terms in the Hamiltonian
of the same dimensionality and symmetry are irrelevant to these
properties.
Such ``universality'' applies, however, only to systems with the
same range of interactions.  Exponents {\it are} affected by long range 
forces which can alter the way correlations diverge at the critical 
temperature.

As a particular
instance of this effect, increasing the
dimensionality can be regarded as the
addition of peculiar long range interactions:  The
$d=2$ Ising model on a $N \times N$ lattice is trivially equivalent
to a ``$d=1$'' model where site $l$ connects to sites $l \pm \hat x$
and $l \pm N \hat x$.  This new model is characterized by a 
coordination number $z=4$ and a regular set of long range bonds.  
The critical exponents were changed by the addition of the 
$l \pm N \hat x$ bonds.
In order to study this effect, and in particular to understand the
importance of the regularity of the additional connections,
we consider in this paper a ``$d=1$'' ferromagnetic Ising model 
\begin{eqnarray}
H=-J\sum_{\langle ij \rangle} S_{i}S_{j} 
\end{eqnarray}
where $S_{i}$ is a classical variable which can take the values
$\pm 1$ and the sum is over ``random'' neighbors 
$\langle i j \rangle$ defined
more precisely as follows. 
Suppose we have
a total of $V$ sites each of which is to have $z$ randomly
chosen neighbors.  Our operational prescription for the construction
of random neighbors is to begin with a list of integers of length $zV$
in which the first
$V$ positive integers are repeated $z$ times.  This list is
randomized by a large number
of pair interchanges.  Sites are defined to be neighbors if they appear
as the $2n-1$ and $2n$ entries in the list, $n=1,2,...{zV \over 2}$.
We eliminate by further randomization
any pairs that appear twice or sites
that are self-connected.
The end result of this procedure is a lattice in which each spin
has $z$ neighbors randomly chosen from the $V-1$ remaining spins\cite{note1}.

For $z=2$ we see that, apart from a trivial relabeling,
we will construct the $d=1$ dimensional
Ising model, if we insist that the lattice have
all sites in the same cluster.  For $z=4$ we would
have a model similar to the $d=2$ Ising Hamiltonian in the sense
that each site would be connected to four neighbors, two of
which could be arbitrarily considered as geometrically close, the other
two of which are not.  However, the disordered nature of the bonds
has an important qualitative effect.  The 
bonds are ``more long ranged'' than the regular d-dimensional
case.  Consider the average separation, 
$\langle \sigma_{ij} \rangle_{c}$,
between two spins, defined as the average over all pairs 
of spins of the minimum number of bonds traversed
in moving from one to another.  (The subscript ``c'' is 
included to emphasize that this is an average over the different
pairs of sites in the lattice rather than the usual thermodynamic
average.  We could also ask about an average over different
possible realizations of the random bonds.  However, we shall assume
that in the thermodynamic limit of large $V$ the lattice is self-averaging.)  
For short-ranged models in any
dimension $d$, the average separation grows with the linear extent $L$.
That is,
$\langle \sigma_{ij} \rangle \approx L = V^{1 \over d}$.  
For the random bond case, $\langle \sigma_{ij} \rangle = a(z) + 
b(z){\rm ln} V$.
The average separation of spins is much smaller.
The role of $z$ is not fundamental in the sense that
it controls only the values of $a$ and $b$, not the logarithmic
dependence on $V$.  This result can be obtained numerically by
explicit construction of lattices and computing
$\sigma_{ij}$.  
In Fig.~1 we show such
numerically obtained values
for our random long range model.

\begin{figure}
\label{lrdist}
\includegraphics[width=3.0in,angle=270]{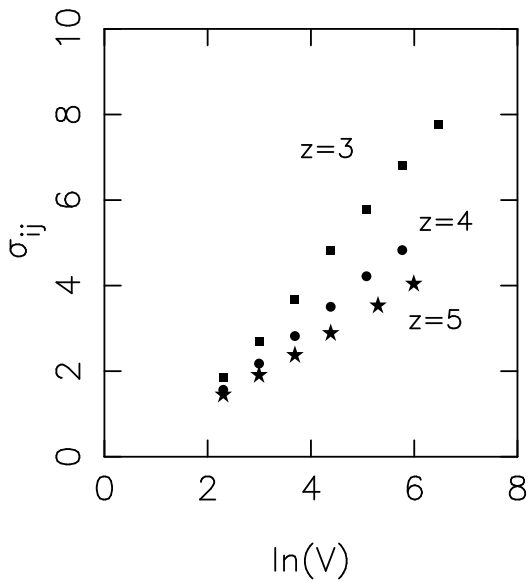}
\caption{
The average separation between spins 
$\langle \sigma_{ij} \rangle_{c}$ is shown as a function of ln$V$ for
different choices of the coordination number $z$.
The slope of the curve is $b(z)$.
}
\end{figure}
          
The behavior of the average separation with volume can also be
motivated analytically by considering a Bethe lattice of coordination
number $z$.
Such a lattice differs from that described above in having no closed loops,
but nevertheless, as we will see in greater detail below, is relevant
to the random bond model we are studying.
The central site from which the lattice is grown has $z$ first neighbors,
$z(z-1)$ second neighbors, and $z(z-1)^{n-1}$ neighbors of distance $n$.
If the lattice is terminated after level $N$, then the total number of sites 
is $V=1+z\sum_{n=1}^{N} (z-1)^{n-1}$ and the distance to the
central site in the thermodynamic limit
is ${1 \over V} z \sum_{n=1}^{N} n (z-1)^{n-1} \approx
{1 \over {\rm ln}(z-1)} {\rm ln} V$.
Thus a Bethe lattice of coordination
number $z$ exhibits a similar logarithmic growth of
$\langle \sigma_{ij} \rangle$ with $V$
and has the particular value $b(z)={1 \over {\rm ln}(z-1)}$.  

This very slow growth, in a fashion qualitatively
independent of $z$ for $z \ge 3$, 
of the ``distance'' between bonds with 
the system size might suggest that the critical properties 
will be more similar to that of the mean field case, where the
distance between all bonds is unity, than the case of finite dimension
where $\langle \sigma_{ij} \rangle \approx V^{{1 \over d}}$.  
Indeed, the apparent absence of a length scale would also suggest mean field
like behavior.  
This is in accordance with the fact that if one defines the dimensionality
of a system\cite{baxter82} as $d={\rm lim}_{n\rightarrow \infty} \,\, 
[{\rm ln} \, c_{n}
/ {\rm ln} \, n] $, where $c_{n}$ is the number of sites within $n$ steps of a
given site, then one finds $d=\infty$.
However, the number of paths between spins is also
relevant in determining the critical behavior\cite{itzhykson89}, and this 
number of paths does depend on $z$, so these arguments suggestive of
mean field behavior are not conclusive.

To make this point more precise, we note that
regular $d$-dimensional lattice Hamiltonians are characterized by the 
existence of short loops of bonds which connect sites back to
themselves.  
In contrast, if the underlying lattice is a Cayley tree, 
sites are connected to neighbors without any such closed loops.
The model discussed here can be viewed as
an alternate way of tying off the ends of a Cayley tree lattice
in which closed loops of size ${\rm ln} V$ are created.
For such structures it is known that the manner in which the boundary,
which is a nonvanishing fraction of the total number of sites
in the thermodynamic limit, is treated is important to the
critical behavior\cite{chen74}.  

Calculation of the energy cost for
introducing a domain wall which divides the lattice in two 
is also instructive.
Consider first the usual Ising model on a hypercubic lattice.
For free boundary conditions in one dimension only a single
bond need be broken.  In two dimensions $V^{{1 \over 2}}$
bonds must be broken, and in $d$ dimensions we need to break
$V^{{d-1 \over d}}$ bonds.  This increasing difficulty of forming a domain
wall is, of course, why the tendency towards ordering is higher as
the dimensionality increases.  In the random bond model we have described,
an attempt to divide the lattice in two equal pieces would on the
average necessitate a breaking of ${V \over 2}$ bonds, a scaling with the
volume $V$ which is the $d \rightarrow \infty$ limit of the usual Ising
case.  This further
points towards a mean field scenario for the critical properties.

We can analyze in a little more detail
the effect of the behavior of $\langle \sigma_{ij} \rangle$ by
considering a high temperature expansion.  To lowest order
\begin{eqnarray}
&\langle& S_{i}S_{j} \rangle \approx \tau^
{\sigma_{ij}}
\nonumber
\\
&\tau&={\rm tanh}(\beta J).
\end{eqnarray}
(The Griffith's inequality in fact guarantees that this is a lower
bound on the correlation function\cite{griffiths72}.)
In 1 dimension for free boundary conditions the shortest path
is the {\it only} path and
this expression is exact.  $\langle S_{i}S_{j} \rangle$
decays exponentially to zero with increasing separation except at $T=T_{c}=0$.

The magnetic susceptibility is a sum over all such correlation functions.
For temperature $T>T_{c}$ where the magnetization is zero,
the susceptibility per site is
\begin{eqnarray}
{1 \over V} \chi &=& {1\over V} \sum_{ij} \langle S_{i}S_{j} \rangle   
\approx V \langle \tau^{\sigma_{ij}} \rangle_{c} 
\nonumber
\\ 
&\approx& V \tau^{\langle \sigma_{ij} \rangle_{c}}  
= \tau^{a(z)}\,\, V^{1- b(z) | {\rm ln}(\tau) | }.
\end{eqnarray}
Thus we estimate a critical temperature at which ${\chi \over V}$  
diverges in the thermodynamic limit as 
\begin{eqnarray}
\tau_{c}={\rm tanh} (\beta_{c} J) \approx e^{-{1 \over b(z) } }.
\end{eqnarray}
This argument, in fact, gives the exact result for the Bethe lattice, 
since if we substitute $b(z)={1 \over {\rm ln}(z-1)}$ into Eq.~(4),
we obtain $\beta_{c}J = {1 \over 2} {\rm ln} {z \over z-1}$,
which is precisely the Bethe lattice critical temperature.
On the other hand, within MFT, ${T_{c} \over J} = z$.  
In Fig.~2 we show, as the full curve, a plot of
${T_{c} \over z}$ vs. $z$ obtained from Eq.~(4) using Bethe lattice values of 
$b(z)$ which are close to those obtained numerically
for the random
bond prescription described above.  The squares are
obtained by scaling monte carlo results to obtain $T_{c}$ as discussed
shortly.

In order to see whether these arguments suggesting some sort of mean field
theory really are valid, 
we have performed monte carlo simulations.
In Fig.~3 we show results for the susceptibility per site,
${\chi \over V}$, vs 
$T$ for $z=3$ and $4$.
The different symbols represent lattices of different size, typically
between $V=200$ and $V=6400$
sites.  The increasing sharpness and height with volume
indicate the presence
of a finite $T_{c}$ which increases with the coordination of the lattice. 
Fig.~4 shows similar results for the specific heat, ${C \over V}$. 
Again, we see an increasingly nonanalytic behavior as $V \rightarrow
\infty$.  The values of $T_{c}$ are consistent with those suggested
by Fig.~3.

\begin{figure}
\label{lrtc}
\includegraphics[width=3.0in,angle=270]{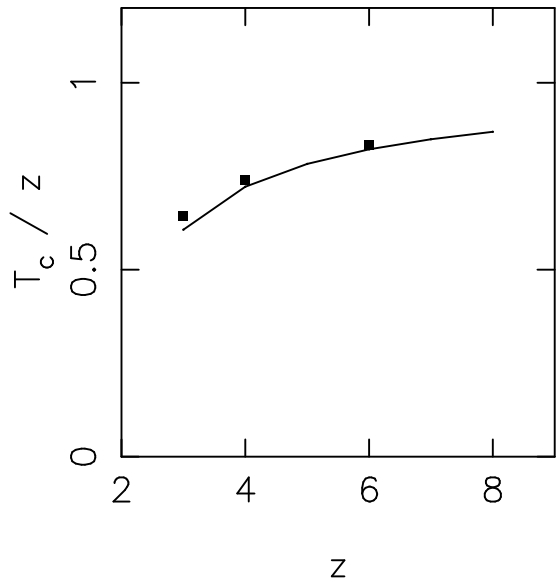}
\caption{
The approximate analytic form for the critical temperature $T_{c}$
obtained from
Eq.~(4)
normalized by the coordination number $z$.
The Bethe lattice form of $b(z)$ has been used.
The squares are values for the normalized critical temperature 
obtained from scaling finite size monte carlo results for the 
susceptibility and the specific heat.
}
\end{figure}
          
To explore the order of the phase transition, we can look at
histograms of the magnetization distribution $P(m)$ for different
temperatures.  At high temperatures $P(m)$ exhibits a 
single peak at $m=0.$  As $T$ is decreased this peak is broadened and
eventually splits into two peaks symmetrically located about $\pm m_{0}$.
At no temperature is evidence seen for the 3 peaked structure characteristic
of a first order transition where $m=0$ and $m \ne 0$ phases are in
coexistence.  This suggests the transition is second order for finite $z$,
as it must be for $z=\infty$.  

\begin{figure}
\label{z3z4chi2}
\includegraphics[width=3.0in,angle=270]{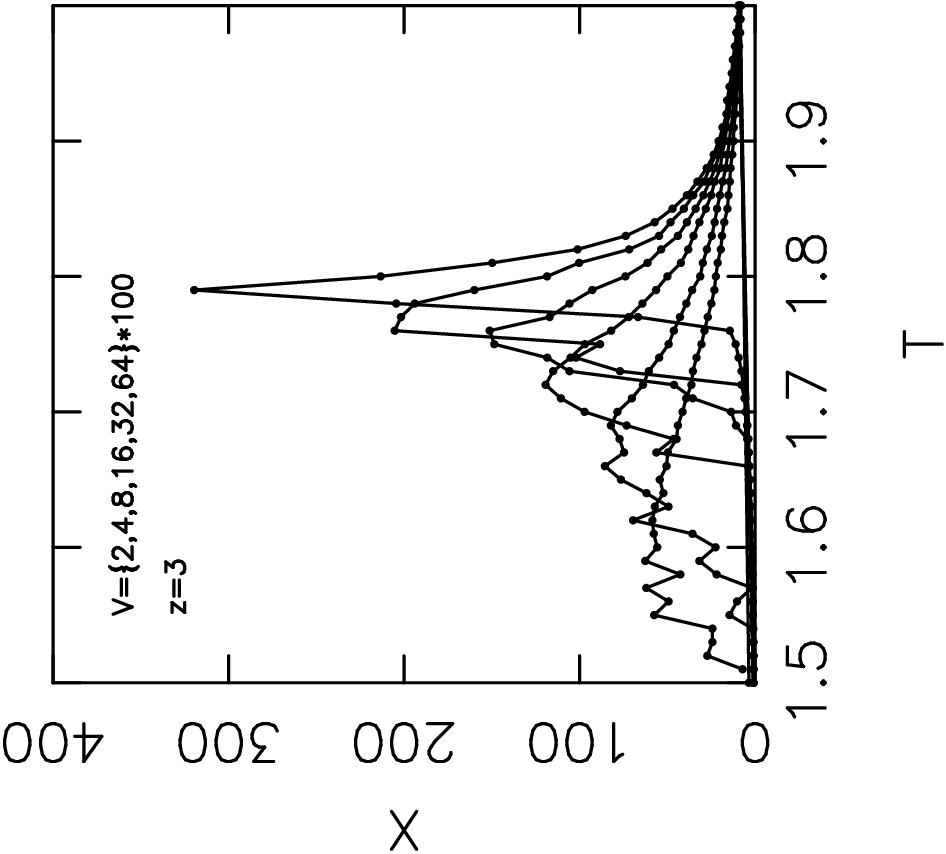}
\includegraphics[width=3.0in,angle=270]{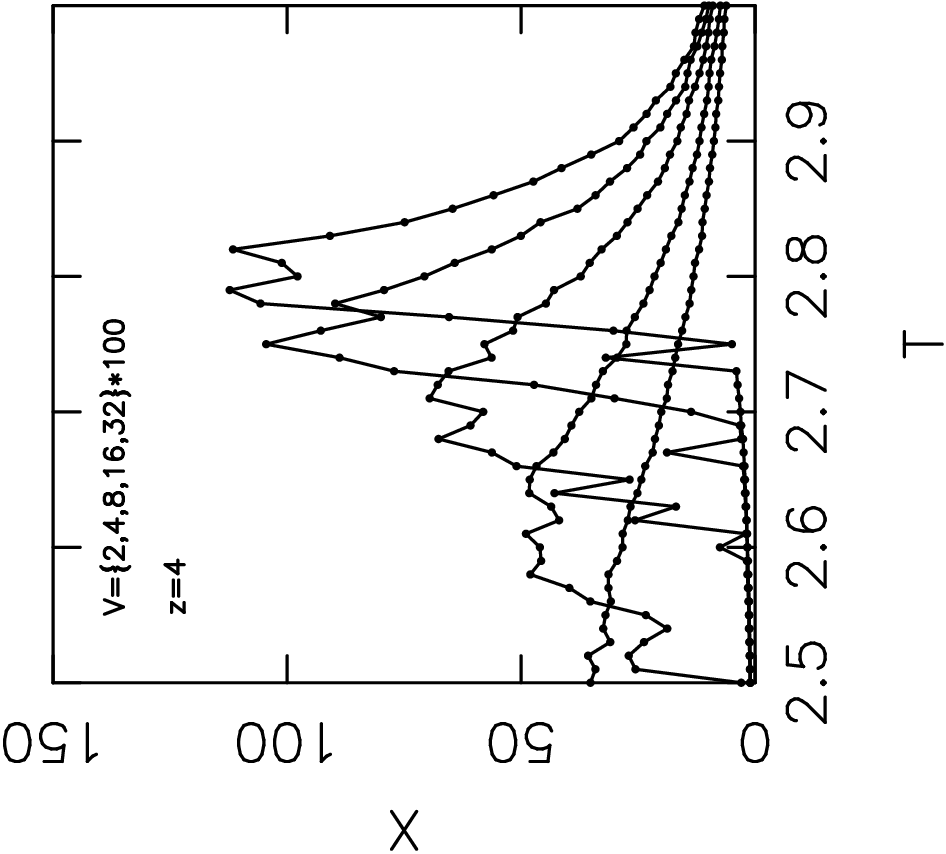}
\caption{
Raw results for the susceptibility per site as a function
of temperature for different sized lattices.  Fig.~3a is for $z=3$
and Fig.~3b for $z=4$.
}
\end{figure}
          
\begin{figure}
\label{z3z4c2}
\includegraphics[width=3.0in,angle=270]{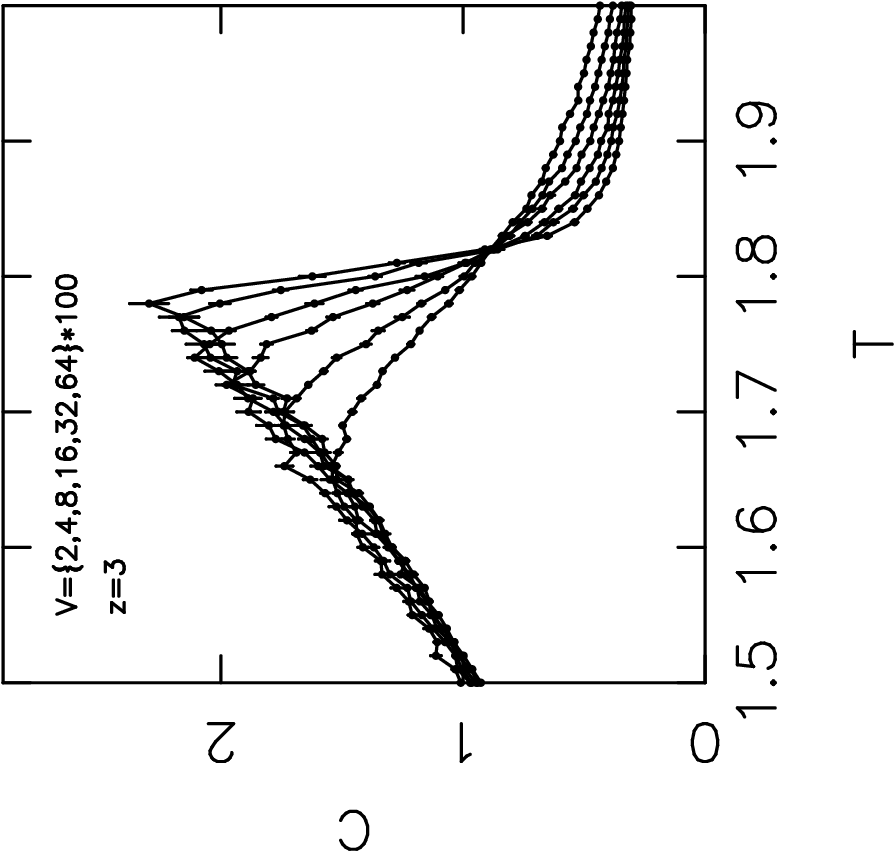}
\includegraphics[width=3.0in,angle=270]{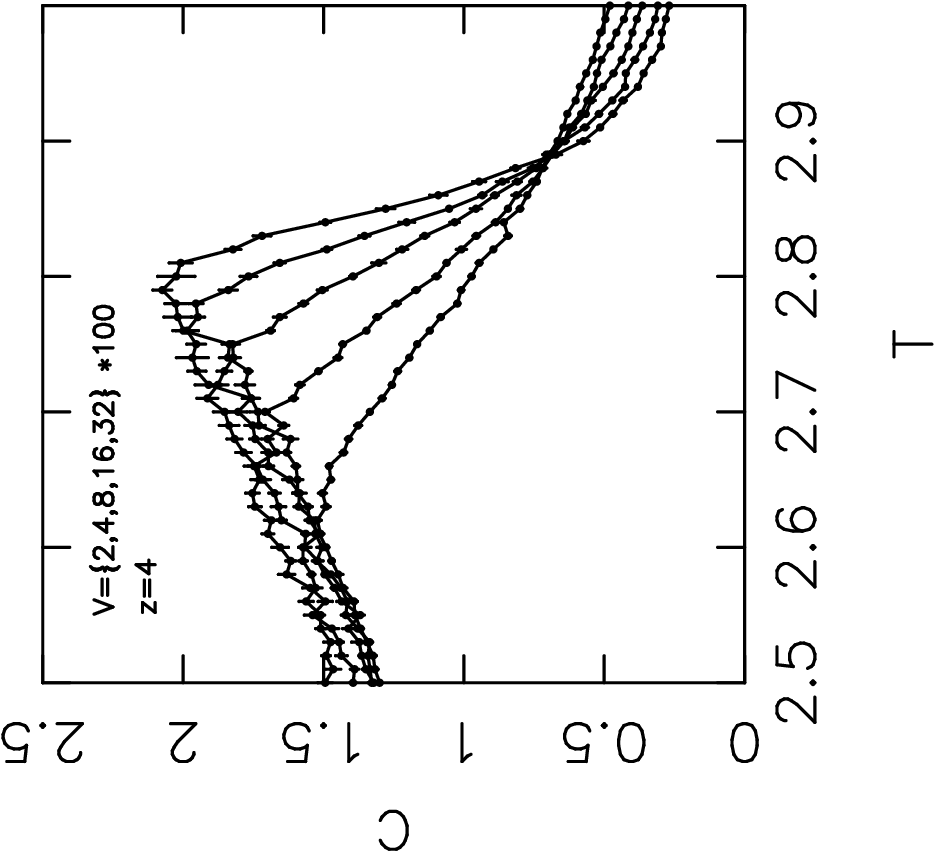}
\caption{
Raw results for the specific heat per site as a function
of temperature for different sized lattices.  Fig.~4a is for $z=3$
and Fig.~4b for $z=4$.
}
\end{figure}
          
It is natural to attempt to perform finite size scaling on the data.
We can first imagine the maxima in the plots of the susceptibility
and specific heat represent values of ``$T_{c}$'' on finite size
lattices.
Conventional scaling theory\cite{barber85}
suggests that
\begin{eqnarray}
T_{c}(V)=T_{c}(\infty)[1-L^{-{1 \over \nu}}].
\end{eqnarray}
We are immediately confronted with the difficulty of relating the linear
extent $L$ to the
number of sites $V$.  Choosing $L=V$ would in the usual d-dimensional 
models simply rescale the exponent $\nu$ by the dimensionality.              
We find that to obtain a reasonable fit to this form we would need to use
an anomalously large value of $\nu$.  In fact, a better $ansatz$ is obtained
by replacing $L$ by ln$V$.
We then find that the best fit for $z=3$
is obtained by $T_{c}(\infty)=1.91 \pm 0.05$
and $\nu=0.52 \pm 0.06$.  
This is, of course, in reasonable agreement 
with the mean field value for the correlation
length exponent $\nu$.                                                     

Turning now to the specific heat $C$ and susceptibility $\chi$,
we expect that curves for different lattices, distinct in their unscaled
forms of Figs.~3 and 4, will fall on universal
curves when scaled with appropriate choices of the 
critical temperature and exponents\cite{barber85}.  That is, for $\chi$
\begin{eqnarray}
{\chi \over L^{ {\gamma \over \nu} } } = f(L^{ {1 \over \nu} } t),
\end{eqnarray}
while for $C$
\begin{eqnarray}
{C \over L^{ {\alpha \over \nu} } } = g(L^{ {1 \over \nu} } t).
\end{eqnarray}
Here $f$ and $g$ are some scaling functions.
In Fig.~5 we show $\chi$
scaled appropriately using $L$=ln$V$ and the critical exponents
$\nu=0.53$ and $\gamma=1.06$.  We have chosen to show
$z=3$, although plots for $z=4$ are similar.
The critical temperature $T_{c}=1.87$.
This scaling plot has somewhat more scatter than the ones obtained
for regular models\cite{binder86}.  However, this can be attributed to the
need to average the results, especially for smaller lattices, over
different configurations of random bonds.  
For the specific heat, we note that the maximum in $C$
in Fig.~4 is insensitive to lattice size.  This translates to a small value
for the exponent $\alpha$, which is zero in the mean field limit.
To estimate the accuracy of these
values of the exponents, we have examined scaling
plots in which the exponents have been shifted.
We see a significant deterioration of the quality 
for changes of more than 10 percent.
                              
\begin{figure}
\label{chiscale}
\includegraphics[width=3.0in,angle=270]{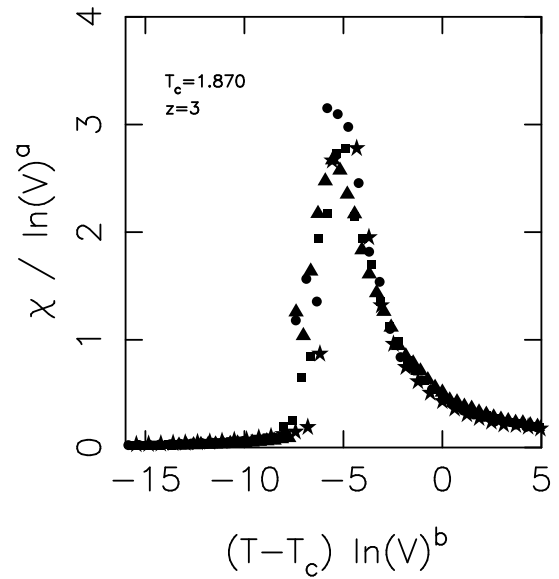}
\caption{
The results of Fig.~3a are shown using the appropriate
scaled forms as described in the text.  The critical exponents used were
$\nu=0.53$ and $\gamma=1.06$.  The critical temperature
$T_{c}$=1.87.
Different symbols correspond to different lattice sizes, $V$=800,
1600, 3200, and 6400.
}
\end{figure}
          
We have seen that the Ising model defined by Eq.~1 and the prescription
for neighbor assignment exhibits mean field like critical behavior
with a length scale which depends logarithmically on the ``volume''
of the system.  
Clearly the regular $d$-dimensional models are realized 
in our model by certain (very)
improbable choices of the random bonds.  We can ask about the
renormalization group flows in the space of possible choices of the 
bonds.  If we construct some quantitative measure of the distance of
a random assignment from the usual regular one, what will be the extent
of the basin of attraction of the fixed points corresponding to the regular
models and their exponents?  Is it vanishingly small, so that in some sense
{\it all} models of this sort flow towards mean field theory, or will there be
some finite region characterized by the regular exponents?

Traditionally, one route to mean field theory is to
increase the
dimensionality of the system, which has the effect
of enhancing the connectivity of the lattice
while leaving the interactions short ranged.
In a similar manner, mean field results are obtained for the critical
properties of the Bethe lattice in the limit where the coordination
$z$ goes to infinity, but again the interaction is
short ranged.  Both these approaches, then, take a route to MFT
via increased numbers of neighbors.  
However, we have seen here a novel realization 
which does not proceed via increased connectivity but rather works with fixed 
coordination and considers a model with extremely long ranged interactions. 
Kac et.al.\cite{kac64} have shown in a one dimensional model 
with an interaction of length scale $R$ that there is a nonzero $T_{c}$
only in the $R \rightarrow \infty$ limit.  There, however, 
$R$ and $z$ are {\it simultaneously} increased.  It is clear from our
work that increasing $R$ alone, at fixed $z$, also induces a MF-like
transition.
Thus it would appear that there are two alternate routes to mean field 
critical behavior:  either diverging $z$ {\it or} diverging $R$ alone
suffice.  As we have pointed out, this is perhaps not too surprising
since neighbors in additional spatial dimensions can equivalently 
be regarded as regular long range bonds.

\noindent
\underbar{Acknowledgements:}
I would like to thank Rajiv R.P. Singh for useful conversations.


\begin{references}

\bibitem{note1}  It is to be emphasized that the connections between sites
$i$ and $j$ are symmetric ones, i.e. once it is determined that two
sites are joined by a bond, then the Hamiltonian Eq.~(1) describes the
thermodynamics in the usual way.  In contrast, within the context of
studies of Kauffman genetic networks, Derrida et.al. have considered
the dynamic evolution of systems of Ising spins with asymmetric connections
in which spin $i$ might influence the subsequent value of $j$, while
$j$ has no direct effect on $i$.  They find that $z=2$ is
a critical value of the connectivity in much the same way that in the
present equilibrium statistical mechanical model there is no phase
transition unless $z>2$.  B. Derrida and Y. Pomeau, Europhys. Lett.
{\bf 1}, 45 (1986).  Further discussion of networks with
dilute, asymmetric bonds is contained in K.E. Kurten, Phys. Lett. {\bf A129},
157 (1989) and J. Physique (France) {\bf 51}, 1585 (1990).

\bibitem{baxter82}  R.J. Baxter, ``Exactly Solved Models in Statistical
Mechanics,'' Academic Press, New York, 1982.

\bibitem{itzykson89}  C. Itzykson and J-M Drouffe in ``Statistical Field Theory,''
Cambridge University Press, New York, 1989.

\bibitem{chen74}  M.S. Chen, J.L. Bonner, and J. Nagle, J. Chem. Phys. {\bf 60},
405 (1974).

\bibitem{griffiths72}  R.B. Griffiths in ``Phase Transitions and Critical Phenomena,
Vol I,''  C. Domb and M.S. Green (eds), Academic Press, New York, 1972.

\bibitem{barber85}  M.N. Barber in ``Phase Transitions and Critical Phenomena,
Vol X,''  C. Domb and J.L. Lebowitz (eds), Academic Press, New York, 1985.

\bibitem{binder86}  K. Binder in ``Monte Carlo Methods in Statistical Physics,''
K. Binder (ed), Springer, New York (1986).

\bibitem{kac64}  M. Kac, G.E. Uhlenbeck, and P.L. Hemmer, J. Math. Phys. {\bf 4},
216 (1964), and J. Math. Phys. {\bf 5}, 60 (1964).

\end{references}
\end{document}